# Using a Cognitive Architecture to consider antiblackness in design and development of AI systems

**Christopher L. Dancy (cdancy@psu.edu)**
Department of Industrial and Manufacturing Engineering & Department of Computer Science and Engineering,
The Pennsylvania State University
310 Leonhard Building, University Park, PA USA

## Abstract

How might we use cognitive modeling to consider the ways in which antiblackness, and racism more broadly, impact the design and development of AI systems? We provide a discussion and an example towards an answer to this question. We use the ACT-R/Φ cognitive architecture and an existing knowledge graph system, ConceptNet, to consider this question not only from a cognitive and sociocultural perspective, but also from a physiological perspective. In addition to using a cognitive modeling as a means to explore how antiblackness may manifest in the design and development of AI systems (particularly from a software engineering perspective), we also introduce connections between antiblackness, *the Human*, and computational cognitive modeling. We argue that the typical eschewing of sociocultural processes and knowledge structures in cognitive architectures and cognitive modeling implicitly furthers a *colorblind* approach to cognitive modeling and hides sociocultural context that is always present in human behavior and affects cognitive processes.

## Introduction

How might we use cognitive modeling to consider the ways in which antiblackness[1], and racism more broadly, impact the design and development of AI systems? There has been a recent surge in scholarship approaching topics such as fairness, ethics, and equity in AI systems (e.g., see *AI, Ethics, and Society* or *Fairness, Accountability, and Transparency*, two conferences that were formed in recent years and focus on those topics.) Approaches in this space tend to focus on fairness and equity in the AI system itself, with solutions that detail ways to modify or test the AI system for forms of fairness (see Leben, 2020, for a discussion of *types* of fairness).

However, the current literature mostly fails to adequately consider two other important pieces of the equation:

- The person (or people) designing, developing, and/or deploying the AI system in question (especially from a *cognitive process* perspective)
- Sociocultural structures and institutions that mediate the way the AI system behaves and learns within an environment. These same structures and institutions also mediate the individuals behind those systems.

When thinking about sociocultural systems and structures, we are pointing particularly to representations of the *Human* (Wynter, 2003; Wynter & McKittrick, 2015). Wynter (2003) traces how representations of the Human (that is, who is considered human and who is considered *other-than* human or *human-Other*) and how the dominant (socioculturally defined) modes of hierarchy that help define the Human have changed throughout recent history. Thinking in-terms of design and development, knowledge structures and representations used to design, develop, and deploy systems (and used by those systems to adapt or *learn*) exist within and re-present sociocultural contexts that designate some as *Other*.

Computational cognitive modeling offers an opportunity to develop computational accounts of the processes that lead to the creation and deployment of AI systems and related computational systems. Ritter (2019) makes a related point in their discussion of applications of cognitive modeling to the system design process. Though they particularly discuss the potential use in a spiral system development process (Pew & Mavor, 2007), the points generally apply to other development processes, especially those one might use in developing software (e.g., the Spiral model), often used to develop and deploy AI systems.

Beyond the typical cognitive models discussed by Ritter (2019), using computational cognitive modeling that includes physiological processes (e.g., Dancy, 2021) and those that include considerations of social processes (e.g., Orr et al., 2019) gives the representational space to consider, quantitatively and qualitatively, how social, cognitive, and physiological processes interact. The ability to understand the realistic interaction between these systems and their effect on behavior becomes especially important for questions related fairness, justice, and equity in the design, development, and deployment of AI systems (see Dancy & Saucier, 2022 for a related discussion of some of these questions that one should consider).

Orr et al. (2019) argues that cognitive architectures (and other systems in the "cognitive levels of scale") should be leveraged in concert with conceptual structures (and dynamics) at the "social level of scale" to develop more complete simulations of human behavior with greater resolution. This paradigm of behavioral simulation, which they call the "Reciprocal Constraints Paradigm" (Orr et al., 2019), calls for cognitive agents to simulate social systems and abstract neurophysiology (upward constraints). The social systems should then constrain cognitive agents, while those cognitive agents constrain the interpretation of neurophysiological behavior (downward constraints).

[1] Here, antiblackness refers to anti-Black racism. For more on connections between AI systems and antiblackness (albeit from a perspective sans cognitive-modeling), see Dancy and Saucier (2022)

Though this work differs in the representations that characterize the upward and downward constraints, there are similarities in recognizing the importance of physiological and social considerations in cognitive and behavioral processes. That is, despite the different *time scales* or *bands* (Newell, 1990) these processes have reciprocal effects albeit ones that differ depending on the scale of behavior. Given that we focus on cognitive models in the development process for this work, we focus the *rational band* of time through a cognitive model lens. The work spans both the cognitive and rational band in terms of generative model simulation (i.e., Ritter, 2019) considered here, but the representational space of the work is one that touches all four of the *biological, cognitive, rational, and social*.

We use the ACT-R/Φ hybrid cognitive architecture and the ConceptNet knowledge graph to consider computational representations that can be used to develop process models that span these bands. In the next sections, we provide more detail of these representations in rational and social bands that we use so that a cognitive model can be used to understand the effects of antiblackness on the design and development process (with a focus on a particular software engineering process). For more detail on the representations in the *biological* and *cognitive* band see (Dancy, 2021) for an understanding of the physio-affective and physio-cognitive connections, and see (Anderson, 2007; Ritter et al., 2019) for a more detailed look at the cognitive process representations.

## *Rational Band* Representations

Thinking through the design and development of AI systems at the *rational band*, is perhaps, the most natural fit for inquiry that aligns with cognitive modeling within cognitive architectures for the task of understanding. It is at this band that we start to think about behavior from the perspective of "knowledge-level" systems (e.g., see Newell, 1990 and also Lieto et al., 2018), at behaviors that span minutes to hours. Here, it is useful to use existing practices in design and engineering (particularly software engineering) as a guide for understanding the cognitive processes enacted within this space of time. We use a Software Engineering framework (Scrum, Schwaber & Sutherland, 2020) to think through the knowledge potentially used during AI system design and development. We also use work that connects processes from Data collection and use, AI development, and Software Engineering (Hutchinson et al., 2021) to move towards an understanding of design and development at this level.

In thinking through the knowledge that is used and enacted during the design and development of AI systems within the rational band of behavior, it's useful to consider the engineering framework that might be used to organize the development behavior and activities. Given the general popularity of agile methodologies and particularly Scrum, we use Scrum to think though behaviors within the scale of minutes to hours. Though Scrum can be thought of from the perspectives of social band as well, our considerations here are the behaviors that span minutes to hours (e.g., development of the product backlog, related agile practices such as the development of user-stories, or development of the system itself).

Hutchinson et al. (2021) argue that datasets used in AI and ML systems are a form of *technical infrastructure* and thus are produced by "goal-driven engineering" processes. Their discussion of Dataset development and curation as an engineering process becomes particularly useful in connecting their discussion of [AI and] ML datasets to considering the cognitive processing of the developer(s). Hutchinson et al. (2021) discuss datasets as forms of engineering models that represent "facts about the world that cannot be experienced directly, nor often replicated". These datasets are often collections of existing digital data, and thus pulled from existing digital knowledge infrastructure. Thus, one can think of these systems as providing a knowledge-level representation (model) of the knowledge used by a person to enact actions within rational (and cognitive) time-scales; relatedly, see (Sparrow et al., 2011) for a discussion on the increased importance of digital computational systems for human knowledge use. Thus, the use of some of these datasets can be extended beyond traditional AI/ML (e.g., Reinforcement learning, or Neural Network-based systems) systems, to generative cognitive models built within cognitive architectures and this may be warranted because these datasets can be thought of as a model of the (extended) knowledge systems used by humans to determine behavior, especially within the rational band. Using these datasets as a model of the knowledge used by designers and developers during cognitive processing and behavior within the *rational band* presents an opportunity to develop models that simulate multi-scale (or in this case, multi-band) representations. While we might use software engineering and design cognition perspectives to develop the task-focused procedural and declarative knowledge for a relevant cognitive model, some datasets can provide a useful *engineering infrastructure* for wider considerations of sociocultural knowledge (e.g., those knowledge structures that encode power structures and hierarchies) with those *task-oriented* procedures and knowledge.

## *Social Band* Representations

Understanding how existing social structures mediate behavior at the individual level is important for understanding contextualized behavior across time. In addition to physical structures and the affordances those structures may provide, it is also useful to consider the *knowledge* structures that are more directly related to behaviors in this band and how those may influence cognitive behavior, whether explicitly or implicitly. Given that such knowledge structures can be learned in diffuse ways across larger time-scales (indeed, knowledge and meaning taken from that knowledge can span generations), it is important to understand how this social (and cultural) knowledge might influence behavior during AI design and development. One can contend that a fundamental aspect of sociocultural knowledge is who is and is not seen as a part of the *Human* (Wynter & McKittrick, 2015) and as others have argued (e.g., Benjamin, 2019; Costanza-Chock, 2020; Noble, 2018) our ability to recognize the humanity is important in the way we design systems. The knowledge structures considered in racist hierarchies that perpetuate antiblackness are best thought of at the social band and time-scale because, though the context may change as environments change, these power structures and hierarchies represented in knowledge persist across time and space (McKittrick, 2006; Wynter, 2003).

To explore design and development from this perspective, large representations of digital knowledge (e.g., knowledge graphs) and large models that encode concepts and relations between concepts (e.g., word embeddings) can prove useful. These models can be thought of not only as *technical infrastructure*, but also as *models of the world* (as discussed in the previous section). The interest in the exploration of bias in these language/knowledge models, ultimately leading to a direct comparison to knowledge communicated by people (Caliskan et al., 2017), adds to the evidence that these models may be useful as a model of world, especially social, knowledge. Due to this primary concern of their connection to knowledge at time-scales in the social band, we discuss a particular model here. We use the ConceptNet knowledge graph and API (Speer et al., 2017) towards this aim of using an existing digital computational model of the world to consider antiblackness in design and development of AI systems. The open-source ConceptNet knowledge graph can be used by AI systems to attach *meanings* to words. Though the network itself is most robust in English and likely transfers some biases from English to other languages, it is a multilingual knowledge graph. The ConceptNet knowledge graph combines knowledge from several sources including crowd-sourcing, certain games, and some resources created by experts.

The ConceptNet API contains an integrated system that is a hybrid of several word embeddings and gives values of (among other things) *relatedness* between terms. Similar to previous work on connecting ACT-R to other sources of knowledge outside of the traditional declarative memory representation (e.g., Kelly et al., 2020; Salvucci, 2014), we are proposing to think through and model using a system that can represent declarative facts, but differs from the standard declarative memory system in ACT-R; that is, to use this *model of the world* to consider relations between concepts, how they encode social systems of power (such as those related to antiblackness), and how this might effect behavior during the engineering process (i.e., as discussed in the previous section).

## Considering Antiblackness in the Design and Development Process

Though ConceptNet has been through processes of "de-biasing" (Speer, 2017), this has not necessarily resulted in the removal of representations of antiblackness if one audits the system with a more critical lens (e.g., see Dancy & Saucier, 2022). This "de-biased" representation of antiblackness is particularly interesting given that one can use the system to compare effects on computational cognitive models across versions (or perhaps, thinking from the human developer perspective, we can look at before and after bias training.) Thus, as also argued by Dancy and Saucier (2022), there exists an opportunity to think beyond just representation and bias by using this model of knowledge about the world and a cognitive model built within a cognitive architecture. We can begin to generate and better understand some ways that the infra-human (and other related racist ideas and concepts) may creep into decision-making. This is not to say that one can solely use these tools to explore antiblackness in AI design and development, but that they can serve as a complement to existing historical and sociocultural perspectives. Computational cognitive models can be used to help explore and probe the artifacts that digitize existing power structures that have produced (and continue to produce) these racist ideas, which are then consumed in a racist bootstrapping of knowledge and action. We also should emphasize that even if we are to move towards a potential process-based explanation of antiblackness at the cognitive and rational level, this does not relinquish the responsibility and agency of individuals and the groups that individual agents form; indeed, concepts such as ethical cognition (Bostrom & Yudkowsky, 2014) and racial literacy (Daniels et al., 2019) must remain an explicit goal even in the face of understanding the mediating cognitive processes.

### Interpreting relatedness in ConceptNet from a *cognitive* and *rational* perspective using ACT-R

Dancy and Saucier (2022) details the ways in which, despite debiasing processes, the system still shows problematic relatedness calculations between racialized concepts and particularly negative representations. As an example, when looking at relatedness between concepts related to humanity (or the lack of it) and racialized "man" (i.e., "black_man", "white_man"), the authors found "black_man" to have a higher relatedness to terms such as "savage", "beast", and "inhuman". Racialized "woman" concepts (again, "black_woman" and "white_woman") are problematic, but (somewhat expectedly) in a different way. While "white_woman" shows almost an exact match in relatedness as "woman", thereby making "woman" and "white_woman" interchangeable, "black_woman" is absent from the system (i.e., black_woman is not a term in the whole knowledge_graph and so relatedness is determined solely by a different algorithm than for the other concepts). As discussed below, these representations are further problematic when one considers the number of edges between racialized concepts and other concepts within the knowledge graph.

From an ACT-R/Phi perspective, these differences in ConceptNet term relatedness values (and indeed edges between terms) are important when considering how a person (or cognitive model) may act given different situations. The relatedness can, essentially, be seen as an important component in a calculation of association strength. In a cognitive-process scenario (i.e., one which involves a typical affective and physiological state), this type of relatedness between terms may be important for Instance-based learning Gonzalez et al., 2003, as well as prospective memory and goal selection Altmann & Trafton, 2002, in decision-making (and also see Thomson et al., 2015). Furthermore, when combining these theoretical perspectives with more realistic physiological and affective variability (e.g., making those same decisions while sleep deprived or stressed), the effects may be multiplied.

Instance-based learning theory describes a feedback loop between retrieving declarative knowledge (*instances*) used to make a decision and the outcome of that decision. The stage of first recognizing the current situation is reliant upon using declarative memory systems. Within ACT-R, this means that the recognition of situations and the knowledge one uses in those situations is guided by the declarative knowledge most available, where the availability of knowledge concepts (typically chunks) is defined by the activation of declarative memory elements (see Anderson, 2007, pp. 91-134; Anderson et al., 2004 for a further discussion of declarative memory activation equations in ACT-R). Thus, a cognitive agent will rely partially on *availability* of potentially competing concepts to ultimately make a decision. Both the subsymbolic role of declarative memory (i.e., being driven by *activation* of a concept) and the symbolic role in making a decision mean that we not only may implicitly retrieve concepts related to human (or less-than human) capacities for understanding how we treat a representation of *black_man*, but also that we may explicitly use these concepts to justify the decision to treat Black people as less-than (e.g., see Fincher et al., 2018). Relatedly, the availability of declarative memory (for our current example, the relatedness/similarity that ultimately affects declarative memory activation) also affects the choice of which goal to pursue (Altmann & Trafton, 2002). The potential goals (and thus problem space explored) by any cognitive agent will be limited by the ontological space that defines their concepts. Being more related to a brute, creature, or beast fundamentally changes the knowledge available for action, as well as the knowledge that will be used to justify and condone action; such knowledge relations help maintain an anti-Black space. Given that, similar to arguments made by Simon (1996) one can think of designing AI systems as being reliant on a designer and developer deciding the goal of the system, it's inner environment (including the technical infrastructure used to train a system and define it's state-action space), and it's outer environment (which the developer is often tasked with modeling or finding a model for and can be related to the models used to train the inner environment), these conceptual relations become problematic even before considering a typical software engineering framework that will help to organize and guide such development.

Though the relatedness for both *man* and *woman* are high even for several of the *less-than*-human terms, this would prove less important for availability in most cases. This expectation stems from the *fan effect* (Anderson & Reder, 1999), which would signal that the large number of edges connected to *man* and *woman* means that it would be less potent in being used as a cue for other concepts and it is more likely that the more specific category (e.g., *black_man*) would be applicable to many situations. Thus, the relatedness of *man* is less material to *black_man* and in the case where *black_man* is directly used as a concept, there is a stark contrast between the relatedness values for the *human* terms and the *less-than-human* terms. The equality between *woman* and *white_woman* (in terms of relatedness) only means that the *woman* does not need to act as a cue to reach the same conclusions (as it appears the edge relations are such that *woman* has a heavy influence on the relatedness of *white_woman*). This discussion says nothing for *black_woman*, which must be extrapolated from other data as the concept is not connected to any other concepts in ConceptNet, not even *woman*; this signals the importance of bringing in *intersectional* (Collins, 2015; Crenshaw, 1989) analysis when understanding these systems.

These availability considerations from ConceptNet as a world knowledge model are intensified when one considers *non-ideal* physiological and affective states. Changes in affect and stress lead to differences in both declarative and procedural memory availability and selection (Dancy, 2021; Schwabe & Wolf, 2013) and can facilitate the switch between using more implicit memory strategies to guide decision-making and action selection (Schwabe et al., 2009). Thus, the implications of a less-than human ontological space are worsened by the fact that we can switch to more implicit memory strategies when under certain states, creating a higher potential to use the biased conceptual knowledge we've received from our environment. That is, any de-biasing attempts we might see in the form of training related to "diversity" or "equity", the developers are likely to be influenced by the dynamics of physiology and affect; most notably for our current purposes, those dynamics associated with stress. This becomes a practical issue for engineers and designers of AI systems as they are not likely to create these intelligent artifacts in a vacuum and under a perfect state, but very realistically while experiencing normal life stressors. Thus, without critically addressing these issues of anti-Blackness constantly and explicitly, we ensure the continuation of a cycle with a new justification.

Considering these results from the perspective of software engineering (particularly a Scrum/agile process for our purposes), we can think how this may affect the Scrum artifacts created. If a member of the team is creating *user-stories* (e.g., with the template of "As a <*user*>, I want <*to perform something*>, so that <*I can achieve some goal*>") with these types of subsymbolic connections (not to say anything for explicit symbolic connections), this will shape the formation of these user stories. Typically, these user stories will then be the artifacts that make-up the product backlog, which is used to designate tasks within each *sprint*. Thus, these artifacts, which play a big role in design and development of a software (AI) system, will be heavily influenced by the knowledge of the developer, who themselves will have an internal memory/knowledge environment that represents an existing (social-scale) system of power and racial hierarchy. This is nothing to say of the other parts of meetings (e.g., daily scrums and sprint planning, reviews, and retrospectives) which themselves may result in problematic changes in development. Fig. 1 gives a high-level picture of how questions, choices, and artifacts created during processes within the Scrum framework will ultimately be  will be mediated not by a cognitive system, which itself is influenced by the existing knowledge and also by stressors (partially mediated by physio-affective processes) experienced during this information processing.

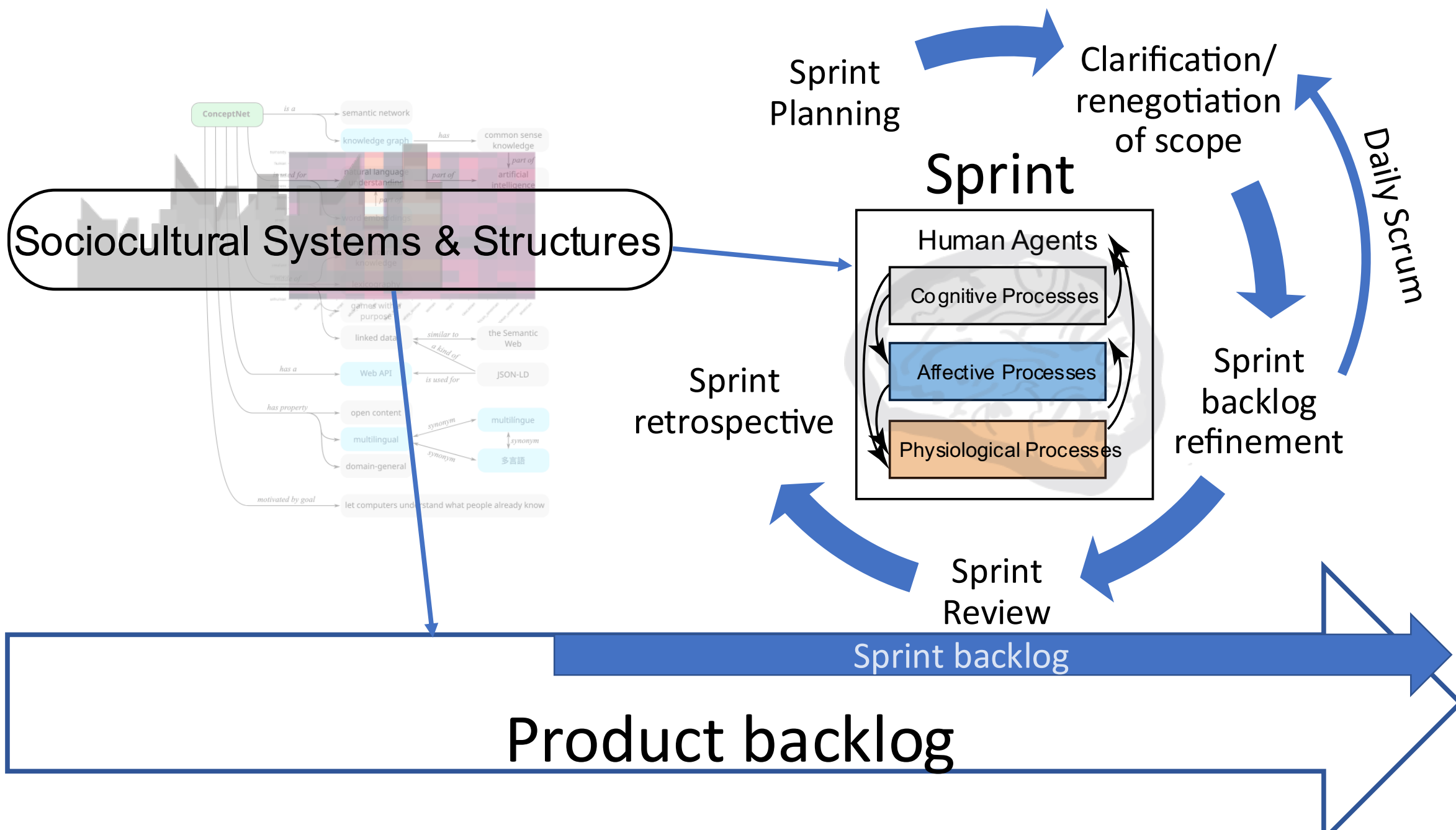

Figure 1. Sociocultural structures influence the Scrum process both through effects on team members during sprints, and through influence of peripheral development of the product backlog.

This problem is further complicated when considering that, despite any de-biasing attempts we might see in the form of training related to "diversity" or "equity", the developers are likely to be influenced by the dynamics of physiology and affect; perhaps most notably for our current purposes, those dynamics associated with stress.

## "Look, a Negro" or Taking into Account the *Sociogenic Principle*.

Fanon (2008), the source of the quote in the section title, in discussing the experience of antiblackness in western contexts, and how fundamental, partially social definitions of what it means to be *human* or *other* influence those placed in either categories within western sociocultural contexts, coins the term sociogeny. He puts forth this concept in addition to phylogeny and ontogeny as an additional layer that determines what it means to be human. Carrying this idea and argument forward, Wynter (2001) adapts the term to sociogenic principle. Wynter uses the sociogenic principle to theorize hybrid "nature/culture" modes of being human; Wynter and McKittrick (2015), and Wynter (2003), trace more recent (western) dominant modes of the Human through history. The sociogenic principle gives us an opportunity to seriously consider how our definitions of the cognitive architecture, or at least the treatment of architectures and cognitive models, may or may not encode fundamental, sociocultural specific aspects of human. As discussed by Wynter (2001), sociocultural knowledge (that operates at the timescales in the social-band) will have foundational effects on behavior causing a "sociocultural situation" to activate a "specific biochemical…correlate". Critically to our use of language-related models as digital models of the world, Wynter (2001) also links the sociocultural mode of the Human to language, particularly the "historico-raical schemas" which are elaborated through a "thousand anecdotes" (and also see a Dancy & Saucier, 2022 for a related discussion relating Fanon's treatment of sociogeny, language, and computational models like ConceptNet).

Nonetheless, the consistent effects of dominant sociocultural knowledge systems (especially those encoding systems of power and oppression such as race) have largely remained hidden and under explored, because cognitive architectures and cognitive models have tended to focus on behavior at the cognitive band of time (though simulating behaviors in the rational band)[2]. Systems and Models that encode world knowledge, such as ConceptNet, give another opportunity to consider how pervasive connections between what it means to be human and race may computationally mediate behavior within the *biological, cognitive, and rational time bands.* In some ways this perspective relates to SGOMS (West & Pronovost, 2009), Orr et al. (2019), and more generally Lieto et al. (2018), but we take aim specifically at racializing hierarchies as a fundamental organizing principle to the social world we operate within. Thus, we are perhaps a level above those in that we are thinking through how to fit (at least) one system of oppression (which is foundational to the current Western context that dominates, but is not exclusive to, the US) within an existing cognitive architecture.

## Conclusion and Future Work

Even with existing computational systems and models of knowledge, there remains work to be done in connecting these systems. In doing this, we seek to avoid multiple models in the rational and social band that ultimately, do not get us closer to understanding how sociocultural knowledge and systems fundamentally organize behavior at lower bands. Lieto et al. (2018) discusses this issue as something related to criticism put forth by Newell (1973), but at the rational band. Thus, in addition to the importance of grappling with our socioculturally contextualized definition of the Human and of the *other* [than human] as laid out by Wynter and McKittrick (2015), there lies an importance in specifying the organizing principles that we will use to develop computational models that span multiple levels.

We plan to continue this work through strengthening connections between ACT-R/Φ and ConceptNet, with an exploration into better ways to combine the declarative memory equations present in ACT-R, the existing ConceptNet knowledge-graph, and the numberbatch system integrated into the ConceptNet API (work such as Salvucci, 2014, is instructive towards this goal). We also plan to explore related existing word embeddings to study how different underlying *technical infrastructure* (i.e., Hutchinson et al., 2021) and methods for determining vectors may affect models developed for the purposes of exploring antiblackness in AI design and development. We also plan to develop computational cognitive models that use world knowledge (starting with ConceptNet) to make decisions during software engineering processes.

Moving to sociocultural processes in models such as ConceptNet, this work would benefit from a more fine-grained analysis of race (i.e., beyond Black and white); using theory posited by Bonilla-Silva (2015) may prove especially useful here. It would also be beneficial to expand beyond race to other sociocultural, power systems that intersect with race such as gender. As discussed in section *Interpreting relatedness in ConceptNet from a cognitive and rational perspective using ACT-R systems*, we have explored some intersections, but more work is needed.

While there has been work in understanding bias in the development of AI systems, cognitive modeling with cognitive architectures has yet to be used to develop a computational process-level understanding of issues in that area. What's more specific focus of antiblackness in design and development, which itself has a "historico-social" context and is structural in ways we must understand, has rarely been explored. Additionally, when social systems have been approached in cognitive modeling, sociocultural systems of power that play a part in our sociocultural definition of the Human have been ignored, resulting in a colorblind approach to modeling. Using cognitive architectures in concert with existing knowledge (including language) models presents a promising method in which to computationally explore antiblackness in the development of AI systems.

[2] This is nothing to say for the ways in which diffuse systems of power and oppression are so foundational to sociocultural knowledge and behavior that they affect not only the perpetual assumed decontextualization of behavior in studying it, but also *who* has an opportunity to study it.